\documentclass[12pt]{article}
\begin{document}
\vspace*{-.6in}
\thispagestyle{empty}
\begin{flushright}
CALT-68-2300\\ 
CITUSC/00-060\\
hep-th/0011078
\end{flushright}
\baselineskip = 18pt

\vspace{.5in}
{\Large
\begin{center}
String Theory Origins of Supersymmetry\footnote{Work
supported in part by the U.S. Dept. of Energy under Grant No.
DE-FG03-92-ER40701.}
\end{center}}

\begin{center}
John H. Schwarz\\
\emph{California Institute of Technology, Pasadena, CA  91125, USA\\
and\\
Caltech-USC Center for Theoretical Physics\\
University of Southern California, Los Angeles, CA 90089, USA}
\end{center}
\vspace{1in}

\begin{center}
\textbf{Abstract}
\end{center}
\begin{quotation}
\noindent The string theory introduced in early 1971 by Ramond,
Neveu, and myself has two-dimensional world-sheet supersymmetry.
This theory,  developed at about the same time that Golfand and
Likhtman constructed the four-dimensional super-Poincar\'e algebra,
motivated Wess and Zumino to construct supersymmetric
field theories in four dimensions. Gliozzi, Scherk, and
Olive conjectured the spacetime supersymmetry of the string theory
in 1976, a fact that was proved five years later by Green and
myself.
\end{quotation}

\vfil \centerline{Presented at the Conference \it 30 Years of
Supersymmetry}

\newpage


\pagenumbering{arabic}

\section{S-Matrix Theory, Duality, and the Bootstrap}

In the late 1960s there were two parallel trends in particle
physics. On the one hand, many hadron resonances were discovered,
making it quite clear that hadrons are not elementary particles.
In fact, they were found, to good approximation, to lie on linear
parallel Regge trajectories, which supported the notion that they
are composite. Moreover, high energy scattering data displayed
Regge asymptotic behavior that could be explained by the
extrapolation of the same Regge trajectories, as well as one with
vacuum quantum numbers called the {\it Pomeron}. This set of
developments was the focus of the S-Matrix Theory community of
theorists. The intellectual leader of this community was Geoffrey
Chew at UC Berkeley. One popular idea espoused by Chew and
followers was ``nuclear democracy" -- that all hadrons can be
regarded as being equally fundamental. A more specific idea was
the ``bootstrap", that the forces arising from hadron exchanges
are responsible for binding the hadrons, as composites of one
another, in a more or less unique self-consistent manner.

The second major trend in the late 1960s grew out of the famous
SLAC experiments on deep inelastic electron scattering. These gave
clear evidence for point-like constituents (quarks and gluons)
inside the proton. This led to Feynman's ``parton'' model, which
was also an active area of research in those days, and eventually to QCD.

String theory, which is the subject I want to focus on here, grew
out of the S-Matrix approach to hadronic physics. The bootstrap
idea got fleshed out in the late 1960s with the  notion of a
duality relating $s$-channel and $t$-channel processes that went
by the name of ``finite energy sum rules'' \cite{Dolen:1967} -
\cite{Harari:1968}. Another influential development was the
introduction of ``duality diagrams'', which keep track of how
quark quantum numbers flow in various processes \cite{Harari:1969,
Rosner:1969}. Later, duality diagrams
would be reinterpreted as string world-sheets, with the quark
lines defining boundaries. A related development that aroused
considerable interest was the observation that the bootstrap idea
requires a density of states that increases exponentially with
mass, and that this implies the existence of a critical
temperature, called the Hagedorn temperature \cite{Hagedorn:1968}
- \cite{Frautschi:1971}.

\section{The Dual Resonance Model}

The bootstrap/duality program got a real shot in the arm in 1968 when
Veneziano found a specific mathematical function that explicitly
exhibits the features that people had been discussing in the
abstract \cite{Veneziano:1968}. The function, an Euler beta
function, was proposed as a phenomenological description
of the reaction $\pi + \omega \to \pi + \pi$ in the narrow
resonance approximation. This was known to be a good
approximation, because near linearity of Regge trajectories
implies that the poles should be close to the real axis. A little
later, Lovelace and Shapiro proposed a similar formula to describe
the reaction $\pi + \pi \to \pi + \pi$ \cite{Lovelace:1968,
Shapiro:1969}. Chan and Paton explained how to incorporate
``isospin'' quantum numbers in accord with the Harari--Rosner
rules \cite{Paton:1969}. Also, within a matter of months Virasoro
found an alternative formula with many of the same duality and
Regge properties that required full $s$-$t$-$u$ symmetry
\cite{Virasoro:1969a}. Later it would be understood that whereas
Veneziano's formula describes scattering of open-string ground
states, Virasoro's describes scattering of closed-string ground
states.

In 1969 several groups independently discovered $N$-particle
generalizations of the Veneziano four-particle amplitude
\cite{Bardakci:1969a} - \cite{Koba:1969b}. The $N$-point
generalization of Virasoro's four-point amplitude was constructed
by Shapiro \cite{Shapiro:1970}. In short order Fubini and
Veneziano, and also Bardakci and Mandelstam, showed that the
Veneziano $N$-particle amplitudes could be consistently factorized
in terms of a spectrum of single-particle states described by an
infinite collection of harmonic oscillators \cite{Fubini:1969a} -
\cite{Fubini:1970}. This was a striking development, because it
suggested that these formulas could be viewed as more than just an
approximate phenomenological descriptions of hadronic scattering.
Rather, they could be regarded as the tree approximation to a
full-fledged quantum theory. I don't think that anyone had
anticipated such a possibility one year earlier. It certainly came
as a surprise to me.

One problem that was immediately apparent was that since the
oscillators transformed as Lorentz vectors, the time components
would give rise to negative-norm ghost states. Everyone knew that
such states would violate unitarity and causality. Virasoro came
to the rescue by identifying an infinite set of subsidiary
conditions, which plausibly could eliminate the negative-norm
states from the spectrum \cite{Virasoro:1970}. These subsidiary
conditions are defined by a set of operators, which form the
famous Virasoro algebra \cite{Fubini:1971}. One price for
eliminating ghosts in the way suggested by Virasoro was that the
leading open-string Regge trajectory had to have unit intercept,
and hence, in addition to a massless vector, it contributes a
tachyonic ground state to the spectrum.

Once it was clear that we were dealing with a system with a rich
spectrum of internal excitations, and not just a bunch of
phenomenological formulas, it was natural to ask for a physical
interpretation. The history of who did what and when is a little
tricky to sort out.  As best I can tell, the right answer -- a
one-dimensional extended object (or ``string'') -- was discovered
independently by three people: Nambu, Susskind, and Nielsen
\cite{Nambu:1970a}--\cite{Fairlie:1970}. The string
interpretation of the dual resonance model was not very
influential in the development of the subject until the appearance
of the 1973 paper by Goddard, Goldstone, Rebbi, and Thorn
\cite{Goddard:1973}. It explained in detail how the string action
could be quantized in light-cone gauge. Subsequently Mandelstam
extended this approach to the interacting theory
\cite{Mandelstam:1973}.

\section{The RNS Model and World-Sheet Supersymmetry}

The original dual resonance model (bosonic string theory),
developed in the period 1968--70, suffered from several unphysical
features: the absence of fermions, the presence of a tachyon, and
the need for 26-dimensional spacetime. These
facts motivated the search
for a more realistic string theory. The first important success
was achieved in January 1971 by Pierre Ramond, who had the
inspiration of constructing a string analog of the Dirac equation
\cite{Ramond:1971b}. A bosonic string $X^{\mu}(\sigma, \tau)$ with
$0\le \sigma\le 2\pi$ has a momentum density $P^{\mu}(\sigma,
\tau) = {\partial\over
\partial\tau}X^{\mu}(\sigma, \tau)$, whose zero mode
        $$p^{\mu} = {1\over 2\pi} \int_0^{2\pi} P^{\mu}
        (\sigma, \tau)~d\sigma$$
is the total momentum of the string.   Ramond suggested
introducing an analogous density $\Gamma^{\mu} (\sigma, \tau )$,
whose zero mode
        $$\gamma^{\mu} = {1\over 2\pi} \int_0^{2\pi}
        \Gamma^{\mu}(\sigma, \tau ) ~d \sigma$$
is the usual Dirac matrix.  He then defined Fourier modes of the
product $\Gamma\cdot P$
        $$F_n = {1\over 2\pi} \int_0^{2\pi} e^{-in\sigma}\Gamma\cdot P ~
        d\sigma. \quad n \in {\bf Z}$$
The zero mode,
        $$F_0 = \gamma \cdot p + {\rm oscillator~terms}$$
is an obvious generalization of the Dirac operator, suggesting a
wave equation of the form
        $$F_0 |{\psi}> = 0$$
for a free fermionic string.\footnote{Ramond included an additional mass parameter,
so that the equation he gave was $(F_0 - m) |{\psi}> = 0$, but it was later shown that
consistency requires $m=0$ \cite{Wu:1973}.}

By postulating the usual commutation relations for $X^{\mu}$ and
$P^{\mu}$, as well as
        $$\{\Gamma^{\mu}(\sigma , \tau), \Gamma^{\nu} (\sigma',\tau)\}
        =4\pi \eta^{\mu\nu}\delta(\sigma - \sigma ' ),$$
he discovered the super-Virasoro (or $N=1$ superconformal) algebra\footnote{Ramond did
not give the central terms.}
        $$
        \{F_m,F_n\} = 2L_{m+n} + {c\over 3}\left(m^2-{1\over 4}\right)
        \delta_{m+n,0}$$
        $$[L_m,F_n ]= \left({m\over 2} - n\right)F_{m+n}$$
        $$[L_m,L_n] = (m-n)L_{m+n} + {c\over 12}(m^3-m)\delta_{m+n,0},$$
extending the well-known Virasoro algebra (given by the $L_n$'s
alone).

At the same time, Neveu and I were working together at Princeton
on the development of a new bosonic string theory containing a
field $H^{\mu}(\sigma, \tau)$ satisfying the same anticommutation
relations as $\Gamma^{\mu} (\sigma , \tau)$, but with boundary conditions
that give rise to half-integral modes. A very similar
super-Virasoro algebra arises, but with half-integrally moded
operators
        $$G_r = {1\over 2\pi} \int_0^{2\pi}e^{-ir\sigma}H\cdot
        P~d\sigma, \quad r \in  {\bf Z} +1/2$$
replacing the $F_n$'s. In our first paper (in February 1971) we
introduced an interacting bosonic string theory based on these
operators \cite{Neveu:1971a}. However, that paper simply appended
additional structure onto the Veneziano model.  One month later,
we presented a better scheme that does not
contain the Veneziano model tachyon at $M^2 = -1$
\cite{Neveu:1971b}. However, it contained a new tachyon at $M^2 =
- 1/2$ that we identified as a slightly misplaced ``pion''. We
thought that our theory came quite close to giving a realistic
description of nonstrange mesons, so we called it the ``dual pion
model.''  This identification arose because only amplitudes with
an even number of pions were nonzero. Thus we could identify a
G-parity quantum number for which the ``pions'' were odd. It was
obvious that one could truncate the theory to the even G-parity
sector, and then it would be tachyon-free. However, we did not
emphasize this fact, because we wanted to keep the pions. Our hope at the time
was that a mechanism could be found that would shift the tachyonic
pion and the massless rho to their desired masses.

In April 1971, Andr\'e Neveu visited Berkeley, where he presented
our results.  He received an enthusiastic reception there because
Bardak\c ci, Halpern, and Mandelstam had all tried previously to
incorporate fermionic fields in string theory.  The person who got
most deeply involved, however, was Mandelstam's student Charles
Thorn.  He, Andr\'e, and I figured out how the $G_r$ operators act
as subsidiary gauge conditions in the
interacting theory \cite{Neveu:1971c}. The key
step was to redefine the vacuum by a `picture-changing' operator.
We called the original string Fock space $F_1$ and the new one
$F_2$.  Only in the $F_2$ picture were the super-Virasoro
conditions realized in a straightforward way.

In the same April-May period we began to appreciate the formal
similarity between our construction and the previous work of
Ramond. This led us to conjecture that our model could be extended
to include Ramond's fermions.\footnote{Neveu's recollection of the sequence of events is
slightly different from mine.}
Neveu and I succeeded in finding a
vertex operator describing the emission of a `pion' from a
fermionic string.  We used it to construct amplitudes for two
fermions and any number of pions \cite{Neveu:1971d}.
Two weeks later Charles Thorn
presented a paper containing the same results \cite{Thorn:1971}.
He also obtained
the first explicit formulas for fermion emission.

Let me now turn to the question of what all this has to do with
supersymmetry.  First of all, it is now understood that the
Virasoro algebra describes two-dimensional conformal
transformations, which can be regarded as analytic mappings of a
Riemann surface.  The infinitesimal generator $L_n \sim
-z^{n+1}{d\over dz}$ corresponds to $z \to z + \epsilon z^{n+1}$.
The super-Virasoro (or superconformal) algebra can be regarded as
a generalization to `super-analytic' mappings of a `super Riemann
surface,' with local coordinates $z$ and $\theta$, where $\theta$
is a Grassmann number.

In August 1971, Gervais and Sakita presented a paper proposing an
interpretation of the various operators in terms of a
two-dimensional world-sheet action principle \cite{Gervais:1971b}.
(See also \cite{Aharonov:1972, Iwasaki:1973} for related work.)
Specifically, they
took the $X^{\mu}(\sigma, \tau)$, which transform as scalars in the world-sheet
theory, together with free Majorana (2-component) fermions
$\psi^{\mu}(\sigma , \tau)$.  The action is
        $$S = {1\over 2\pi} \int d\sigma d\tau \left\{
        \partial_{\alpha}X^{\mu}\partial^{\alpha}X_{\mu}-i
        \overline{\psi}^{\mu}\rho^{\alpha}
        \partial_{\alpha}\psi_{\mu}\right\},$$
where $\partial_{\alpha}$ are world-sheet derivatives $({\partial
\over
\partial \tau},~ {\partial\over \partial\sigma})$ and $\rho^{\alpha}$ are
two-dimensional Dirac matrices.  They noted that this has a global
fermionic symmetry: The action $S$ is invariant under the
supersymmetry transformation
        $$ \delta X^{\mu} = \bar \epsilon\psi^{\mu}$$
 $$ \delta \psi^{\mu} = -i\rho^{\alpha}\epsilon\,
        \partial_{\alpha}X^{\mu},$$
where $\epsilon$ is a constant infinitesimal Majorana spinor. So
this demonstrated that the theory has global world-sheet supersymmetry.
I think that this was the first consistent supersymmetric action to be identified.
However, it did not occur to us at that time to explore whether the corresponding string theory
could also have spacetime supersymmetry. Perhaps the presence of the
tachyonic ``pion'' in the spectrum prevented us from considering the possibility.
A few years later, this
theory was also explored by Zumino \cite{Zumino:1974}, a fact
which I think was historically important in setting the stage for
his subsequent work with Wess \cite{Wess:1974} on supersymmetric
field theory in four dimensions.

When $\psi^{\mu}$ has periodic boundary conditions (and hence
integrally-labeled Fourier modes), or the corresponding open string boundary
conditions, it corresponds to Ramond's
$\Gamma^{\mu}(\sigma , \tau)$.  When it is taken to be
antiperiodic, it corresponds to the $H^{\mu}(\sigma , \tau)$
introduced by Neveu and me. The operators $F_n$ or $G_r$
correspond to Fourier modes of the supersymmetry Noether current,
just as $L_n$ corresponds to modes of the two-dimensional
energy--momentum tensor. A deeper understanding of the
significance of the super-Virasoro gauge conditions became
possible following the development of supergravity. That theory
involved making space-time supersymmetry local, so it became natural
to attempt the same for world-sheet supersymmetry.  This was
achieved by introducing a two-dimensional `zweibein' field
$e_{\alpha}^a$ that describes the geometry of the world sheet and
a Rarita-Schwinger field $\chi_{\alpha}$, which is a gauge field
for world-sheet supersymmetry \cite{Brink:1976, Deser:1976}. The
superconformal symmetry arises as constraint conditions after the local symmetries
are used to choose a covariant gauge (rather
like Gauss's law in electrodynamics).

Having identified the subsidiary constraint
conditions, it became
plausible that one could prove that the spectrum of physical
propagating degrees of freedom is ghost-free. The counting looked
encouraging, because the constraints were in one-to-one
correspondence with the time components of the oscillators
($\alpha_n^0 \leftrightarrow L_n$ and $b_r^0 \leftrightarrow
G_r$). Indeed, inspection of some low-lying levels supported this
conjecture. But it was not known for sure whether this was true
for the entire spectrum.

The proof of the no-ghost theorem began with the construction of a
set of vertex operators that create physical excitations
satisfying the Virasoro constraints by Del Giudice, Di Vecchia,
and Fubini \cite{DelGiudice:1972}. Shortly afterwards, Brower and
Thorn  worked out the algebra of these operators
\cite{Brower:1971} and Brower used these results to prove that the
bosonic string
spectrum is ghost-free for $D\leq26$ \cite{Brower:1972}. (There
are ghosts for $D> 26$.) A somewhat different proof was obtained
by Goddard and Thorn, who also showed that the dual pion model is
ghost-free for $D\leq 10$ \cite{Goddard:1972}. The no-ghost
theorem for the dual pion model was also established using
Brower's methods by myself \cite{Schwarz:1972} as well as by
Brower and Friedman \cite{Brower:1973}. Other related results were subsequently
obtained by Gervais and Sakita \cite{Gervais:1973}, Olive and
Scherk \cite{Olive:1973}, and Corrigan and Goddard
\cite{Corrigan:1974}.

\section{Spacetime Supersymmetry}

In the NS (bosonic) sector the mass formula is $M^2 = N -
\frac{1}{2}$, where
\[
N = \sum_{n > 0} \alpha_{-n} \cdot \alpha_n + \sum_{r>0} r b_{-r}
\cdot b_r,
\]
which is to be compared with the formula $M^2 = N - 1$ of the
bosonic string theory.  This time the number operator $N$ has
contributions from the $b$ oscillators as well as the $\alpha$
oscillators.\footnote{The reason that the normal-ordering constant is
$-1/2$ instead of $-1$ works as follows: Each transverse $\alpha$
oscillator contributes $-1/24$ and each transverse $b$ oscillator
contributes $-1/48$. The result follows since the bosonic theory
has 24 transverse directions and the superstring theory has 8
transverse directions.} Thus the ground state, which has $N = 0$,
is now a tachyon with $M^2 = - 1/2$.

This is where things stood until the 1976 work of Gliozzi, Scherk,
and Olive \cite{GSO}.  They noted that the spectrum admits a
consistent truncation (called the GSO projection), which is
necessary for the consistency of the interacting theory.  In the
NS sector, the GSO projection keeps states with an odd number of
$b$-oscillator excitations, and removes states with an even number
of $b$-oscillator excitations. (This corresponds to projecting onto
the even G-parity sector of the dual pion model.) Once this rule
is implemented the only possible values of $N$ are half integers,
and thus the spectrum of allowed masses is integral
\[
M^2 = 0,1,2, \ldots.
\]
In particular, the bosonic ground state is now massless.  So the
spectrum no longer contains a tachyon. The GSO projection also
acts on the R sector, where there is an analogous restriction on
the $d$ oscillators.  This amounts to imposing a chirality
projection on the spinors.

Let us look at the massless spectrum of the GSO-projected theory.
The ground state boson is now a massless vector, represented by
the state $\zeta_\mu b_{-1/2}^\mu |0;p\rangle$, which
has $d - 2 = 8$ physical polarizations.  The ground state fermion
is a massless Majorana--Weyl fermion which has $\frac{1}{4} \cdot
2^{d/2} = 8$ physical polarizations.  Thus there are an equal
number of massless bosons and fermions, as is required for a theory with
spacetime supersymmetry.  In fact, this is the pair of fields that
enter into ten-dimensional super Yang--Mills theory.  The claim is
that the complete theory now has spacetime supersymmetry.

If there is spacetime supersymmetry, then there should be an equal
number of bosons and fermions at every mass level.  Let us denote
the number of bosonic states with $M^2 = n$ by $d_{NS}(n)$ and the
number of fermionic states with $M^2 = n$ by $d_R(n)$.  Then we
can encode these numbers in generating functions
\[
f_{NS}(w) = \sum_{n=0}^\infty d_{NS} (n) w^n
\]
\[
=\frac{1}{2\sqrt{w}}
\left(\prod_{m=1}^\infty \left(\frac{1+w^{m-1/2}}{1 -
w^m}\right)^8 - \prod_{m=1}^\infty \left(\frac{1 -
w^{m-1/2}}{1-w^m}\right)^8\right)
\]
and
\[
 f_{R}(w) =
\sum_{n=0}^\infty d_R (n) w^n = 8 \prod_{m=1}^\infty
\left(\frac{1+w^m}{1-w^m}\right)^8.
\]
The $8$'s in the exponents refer to the number of transverse directions in ten
dimensions.  The effect of the GSO projection is the subtraction of the second
term in $f_{NS}$ and reduction of coefficient in $f_R$ from 16 to 8.  In 1829,
Jacobi discovered the formula\footnote{He used a different notation, of course.}
\[
f_R (w) = f_{NS} (w).
\]
For him this relation was an obscure curiosity, but we now see
that it provides strong evidence for supersymmetry of the
GSO-projected string theory in ten dimensions.  A complete proof
of supersymmetry for the interacting theory was constructed by
Green and me five years after the GSO paper~\cite{Greena}. We
developed an alternative world-sheet theory to describe the
GSO-projected theory \cite{Green:1984}. This formulation has as
the basic world-sheet fields $X^{\mu}$ and $\theta^a$,
representing ten-dimensional superspace. Thus the formulas can be
interpreted as describing the embedding of the world-sheet in
superspace.

\section{Concluding Remarks}

Compared to the older RNS formulation, The GS formulation has a
number of advantages and disadvantages. The main advantage is that
it makes the spacetime supersymmetry manifest, whereas that
symmetry is an extremely obscure in the RNS formalism, as we have
just seen. A significant disadvantage of the GS formalism is that
it involves a subtle combination of first-class and second-class
constraints that cannot be disentangled covariantly. As a result,
covariant quantization becomes a real problem. Green and I showed
that the quantization is very simple and straightforward in
light-cone gauge and used it to carry various tree and one-loop
computations. However, from a more fundamental point of view, the
lack of a satisfactory covariant quantization procedure is a
serious shortcoming.

There have been many attempts over the years to address the
problem of covariant quantization. As far as I know, all proposals
prior to one this year by Berkovits have severe problems.
Berkovits has introduced a new version of the GS formalism, which
involves the use of a ``pure spinor'' $\lambda$
\cite{Berkovits:2000}. Even though $\lambda$ is a spacetime
Majorana spinor, it is a commuting world-sheet field that
satisfies the constraints $\lambda \gamma^{\mu} \lambda =0$. It is
conceivable that Berkovits proposal will be successful.

In addition to making supersymmetry obscure, the RNS formalism has
a second drawback. Namely, it is not well-suited to incorporating
background fields belonging to the Ramond--Ramond sector. In the
GS formalism, on the other hand, there is no special difficulty
associated to RR backgrounds. This issue becomes important in the
context of the AdS/CFT duality that relates N = 4 gauge theory to
type IIB string theory in an AdS${}_5 \times S^5$ background. Here
the background also includes an RR five-form field strength.
Because of the current inability to handle such backgrounds, most
studies have focused on the supergravity approximation to the
string theory, which is sufficient only in a certain limit.
Various authors have attempted to use the GS formalism to handle
the RR background, but then the problem of quantization arises. I
think it might be possible to overcome this difficulty by using
Berkovits' formalism, if one can figure out how to apply it in
this context. This problem is important, because some modification
of this configuration might lead someday to a scheme that
describes QCD. So, if these issues can be sorted out, the original
dream of string theory -- to compute the properties of hadrons --
could still be realized!

\end{document}